\def\year{2024}\relax
\documentclass[letterpaper]{article} 
\usepackage{aaai24}  
\usepackage{times}  
\usepackage{helvet}  
\usepackage{courier}  
\usepackage[hyphens]{url}  
\usepackage{graphicx} 
\usepackage{soul}
\usepackage{subcaption}
\usepackage{xspace}
\usepackage{amssymb}
\usepackage{booktabs}
\usepackage[T1]{fontenc}

\usepackage{floatrow}
\newfloatcommand{capbtabbox}{table}[][\FBwidth]
\DeclareCaptionLabelFormat{andtable}{#1~#2  \&  \tablename~\thetable}

\usepackage{natbib}  
\usepackage{caption} 
\DeclareCaptionStyle{ruled}{labelfont=normalfont,labelsep=colon,strut=off} 
\usepackage{algorithmic}
\usepackage{xcolor}
\usepackage{amsmath, amsfonts}
\usepackage{dsfont}

\usepackage{newfloat}
\usepackage{listings}
\floatstyle{ruled}
\newfloat{listing}{tb}{lst}{}
\floatname{listing}{Listing}
\usepackage{algorithm2e}

\newcommand{\sysname}{{CHAHAK}}
\newcommand{\fullsysname}{{\textbf{C}ampaign for \textbf{H}ealth-resource \textbf{A}llocation using \textbf{H}istorical data in \textbf{A}I based system for \textbf{K}ilkari}}

\title{

 Improving Health Information Access in the World's Largest Maternal Mobile Health Program via Bandit Algorithms
}
\author{
    Arshika Lalan\textsuperscript{\rm 1}\thanks{Joint first author},
    Shresth Verma\textsuperscript{\rm 1}$^*$,
    Paula Rodriguez Diaz\textsuperscript{\rm 2},
    Panayiotis Danassis\textsuperscript{\rm 2},
    Amrita Mahale\textsuperscript{\rm 3},
    Kumar Madhu Sudan\textsuperscript{\rm 3},
    Aparna Hegde\textsuperscript{\rm 3},
    Milind Tambe\textsuperscript{\rm 1},
    Aparna Taneja\textsuperscript{\rm 1}
 }

\affiliations{
\textsuperscript{\rm 1}Google Research India\\
\textsuperscript{\rm 2}Harvard University\\
\textsuperscript{\rm 3}ARMMAN \\
\{arshikal,vermashresth\}@google.com \{prodriguezdiaz,pdanassis\}@g.harvard.edu \{amrita,madhu,aparnahegde\}@armman.org, \{milindtambe,aparnataneja\}@google.com
}

\usepackage{bibentry}

\begin{document}

\maketitle

\begin{abstract}

Harnessing the wide-spread availability of cell phones,  many nonprofits have launched mobile health (mHealth) programs to deliver information via voice or text to beneficiaries in underserved communities, with maternal and infant health being a key area of such mHealth programs. Unfortunately, dwindling listenership is a major challenge, requiring targeted interventions using limited resources. This paper focuses on Kilkari, the world's largest mHealth program for maternal and child care -- with over 3 million active subscribers at a time --  
launched by
India's Ministry of Health and Family Welfare (MoHFW)
and run by the non-profit ARMMAN. We present a system called \sysname{} that aims to reduce automated dropouts as well as boost engagement with the program through the strategic allocation of interventions to beneficiaries. 
Past work in a similar domain has focused on a much smaller scale mHealth program
and used markovian restless multiarmed bandits  to optimize a single limited intervention resource. However this paper demonstrates the challenges in adopting a markovian approach in Kilkari; therefore  
\sysname{} instead relies on non-markovian time-series restless bandits, and optimizes
multiple interventions to improve listenership. We use real Kilkari data from the Odisha state in India to show \sysname{}'s effectiveness in harnessing multiple interventions to boost listenership, benefiting marginalized communities. When deployed \sysname{} will assist the largest maternal mHealth program to date.  

\end{abstract}

\section*{Introduction}
Ensuring the well-being and health of pregnant women and newborn children is a significant challenge in the developing world \cite{UNWomen}.
According to WHO~\cite{WHO2023}, almost 800 women died everyday in 2020 from pregnancy and childbirth related issues. Many of these deaths are preventable with access to critical health information in a timely manner. 
Mobile Health (mHealth) programs make use of mobile phones and other wireless communication technologies to improve access and raise awareness on healthcare information. ARMMAN (\url{https://armman.org/}) is a non-profit organization in India that conducts several such mHealth programs to improve the access of pregnant women, mothers of infants, and their families to critical preventive care information \cite{mMitra}. One program that ARMMAN runs is \emph{Kilkari}~\cite{KilkariArticle},  the \emph{world's largest maternal mHealth messaging program, and it is the focus of this paper}. Kilkari was launched in 2016 by India's Ministry of Health \& Family Welfare (MoHFW) as a free mHealth education service sending women preventive care information though \emph{Automated Voice Messages (AVMs)} during pregnancy and child infancy. 

A major challenge for mHealth programs (including Kilkari) is that of dwindling engagement over time~\cite{amagai2022challenges,mate2022field}, i.e., beneficiaries stop listening to the AVMs. 
In addressing this challenge, SAHELI~\cite{verma2022saheli} is a key previous deployed system that has focused on minimizing such disengagement by optimizing limited interventions (phone calls by health workers). SAHELI was the first system to deploy bandit algorithms, specifically markovian restless multi-armed bandits, for  allocating limited interventions in  ARMMAN's smaller scale maternal mHealth program, mMitra \cite{mMitra}. mMitra serves approximately 200,000 beneficiaries within Mumbai's urban population. In contrast, Kilkari serves around \emph{3.2 million} beneficiaries at a time across 18 states and union territories in India. Despite the similarities in both being maternal mHealth programs, mMitra and Kilkari have several fundamental differences. Most importantly, the lack of demographic and personal information in Kilkari necessitates the use of novel \emph{non-markovian} Time-Series Bandits (TSB) to plan interventions -- i.e., past SOTA techniques on restless multi-armed bandits are not applicable in this context. Additionally, there is the need to infer time-slot preferences for sending AVMs to beneficiaries, requiring the use of stochastic (not restless) bandits.

We propose to deploy a system called \fullsysname{} or \textbf{\sysname{}} (Hindi for ``chirping of birds'') to boost the engagement of beneficiaries with the mHealth program, and, thus, ultimately positively impact behavioural and health outcomes. To that end, we address two critical optimization challenges.
First, we optimize the appropriate time-slot to send AVMs to each beneficiary. Practical problems such as limited access to phones due to shared family phones for many women, working hours, house chore responsibilities etc.,  significantly affect the likelihood of engagement in a given time-slot. 
Second, we  optimize the limited interventions (e.g., support calls, or home visits from health workers) so as to maximize engagement, given the history of the beneficiaries' behaviour. Thus, our contributions are: 



\begin{enumerate}
    \item \textbf{AVM time-slot optimization.} We develop a UCB-based~\cite{auer2002finite} approach that improves the timing of AVMs according to beneficiaries' listenership patterns. This work is the \emph{first} to optimize such time slot selection.

    \item \textbf{Intervention planning.} In absence of rich demographic features in Kilkari, we leverage beneficiaries' listenership history and model the problem as a non-markovian Time-Series Bandit (TSB) to determine which beneficiaries should receive which intervention (such as a visit by an ASHA worker or an automated voice call reminder), given our limited intervention budget. This work is the first to model multiple inteventions via TSB.

    \item \textbf{Secondary evaluation with data from 4000 real-world beneficiaries.} We use \textit{anonymized} real-world Kilkari data from Odisha state in India to show \sysname{}’s effectiveness in identifying beneficiaries' preferred time slot, and in harnessing multiple interventions to boost listenership.
\end{enumerate}

When deployed, \sysname{} will benefit millions of underserved and marginalized women and children in India. While we focus on Kilkari, the proposed methodological contributions have the potential to \emph{perform well on any other multi-action, non-Markovian bandit setting}.

\section*{Discussion and Related Work} \label{sec: related works}

Adherence to healthcare programs is a well-studied problem in a plethora of domains such as cardiac problems~\cite{son2010application, corotto2013heart}, tuberculosis~\cite{Killian_2019, 10.1001/archinte.1996.00440020063008}, HIV \cite{HIV}, elderly care \cite{pollack2002pearl}, lifestyle changes \cite{liao2020personalized}, and more. These works either perform one-shot prediction of high drop-out risk beneficiaries, or involve sequential decision making where one has to select a subset of beneficiaries to intervene at each time-step given limited resources (number of interventions), in order to maximize adherence. Specifically on the latter, 
state-of-the-art approaches tend to model the sequential allocation problem as a Restless Multi-Armed Bandit (RMAB) problem~\cite{nishtala2020missed, mate2020collapsing,mate2022field}.

SAHELI~\cite{verma2022saheli} is the first deployed application of RMAB in public health, and is in continuous use by ARMMAN for mMitra, the maternal mHealth program operating in the Indian city of Mumbai. Despite a similar setting, there are some notable limitations of SAHELI. First, SAHELI assumes that beneficiaries follow the Markov property, which might not be the case in many applications. In fact, in a section ahead we show that beneficiaries' behaviour in Kilkari is co-related to their past listenership patterns, which violates the Markov assumption. Additionally, SAHELI requires access to a rich set of demographic features which are key to transferring knowledge from historic data to newly enrolled beneficiaries. Instead, \sysname{} only has access to listenership logs. Hence utilizing these is crucial in predicting beneficiaries' behaviour. Moreover, SAHELI relied on live service calls as the only intervention, however there are multiple interventions possible in \sysname{}, with varying impact and budget. Inspired by~\cite{danassis2023TARI}, we introduce a novel Multi-action, Non-Markovian Time Series Bandit Model formulation for the problem.

Finally, another key difference between \sysname{} and past work (e.g.,~\cite{verma2022saheli,danassis2023TARI}) is that \sysname{} attempts to reduce the number of \emph{automatic dropouts}, by optimizing for each beneficiary's preferred time slots for receiving AVMs. 
AVMs are not currently optimized as per beneficiary time-slot preferences due to lack of knowledge of phone ownership, working hours, or any other indicator on time-slot preference~\cite{bashingwa2023can}. This aggravates the problem of drop-offs whereby if a beneficiary has less than 25\% engagement in the past 6 weeks, they are automatically dropped off from the system. This automatic dropoff is imposed due to the scale at which Kilkari operates; serving millions of beneficiaries imposes infrastructural and logistic limitations, and thus critical bandwidth can not be spent on AVMs to incorrectly recorded numbers or ones no longer in service.

\section*{\sysname{} Overview}

Kilkari sends 72 Automated Voice Messages (AVMs), one per week,  on topics such as maternal and child health, immunization, family planning, etc. to pregnant women and new mothers to share critical health information 
starting from the second trimester of pregnancy until the child is one year old. 
To increase the reach of these messages, 
a beneficiary receives up to a maximum of 9 call attempts across 4 days (until they pick up) to maximise the probability of receiving the message for the intended week. 

\begin{figure}
    \centering
    \begin{subfigure}[b]{0.6\textwidth}   {\includegraphics[width=\linewidth,height=3cm]{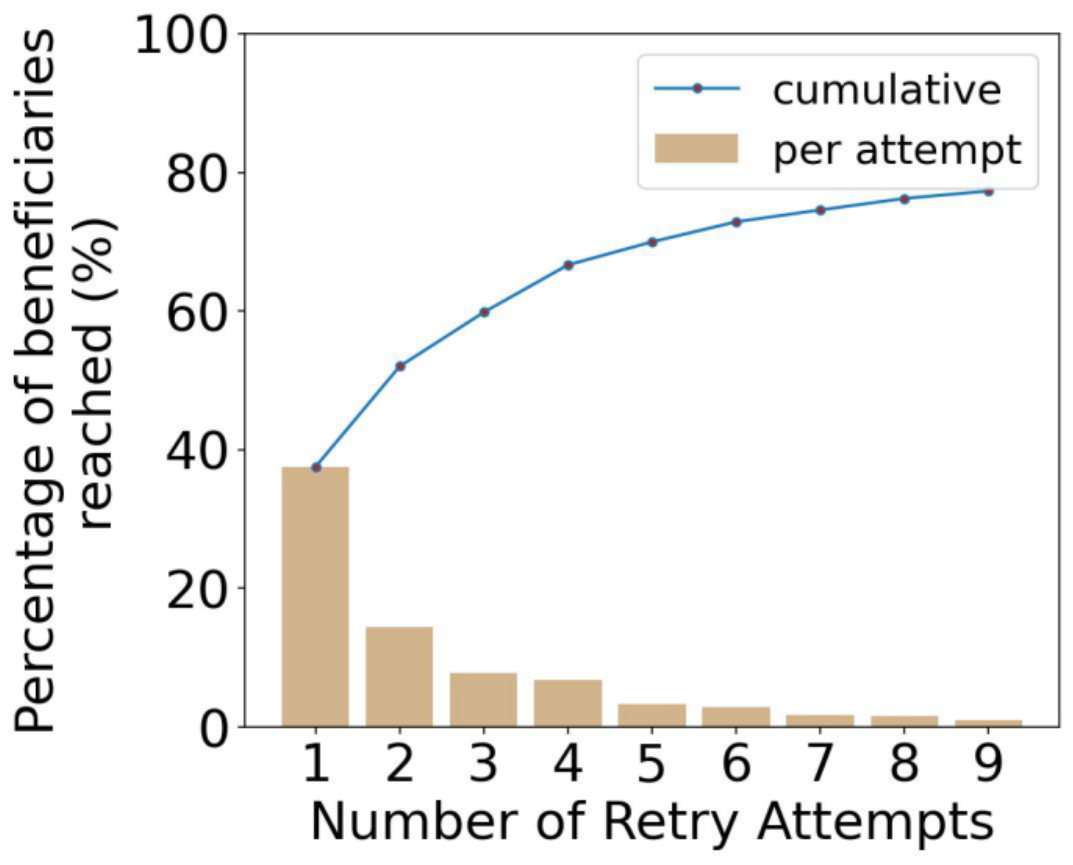}} 
    \end{subfigure}%
    \begin{subfigure}[b]{0.4\textwidth}   {\includegraphics[width=\linewidth,height=3cm]{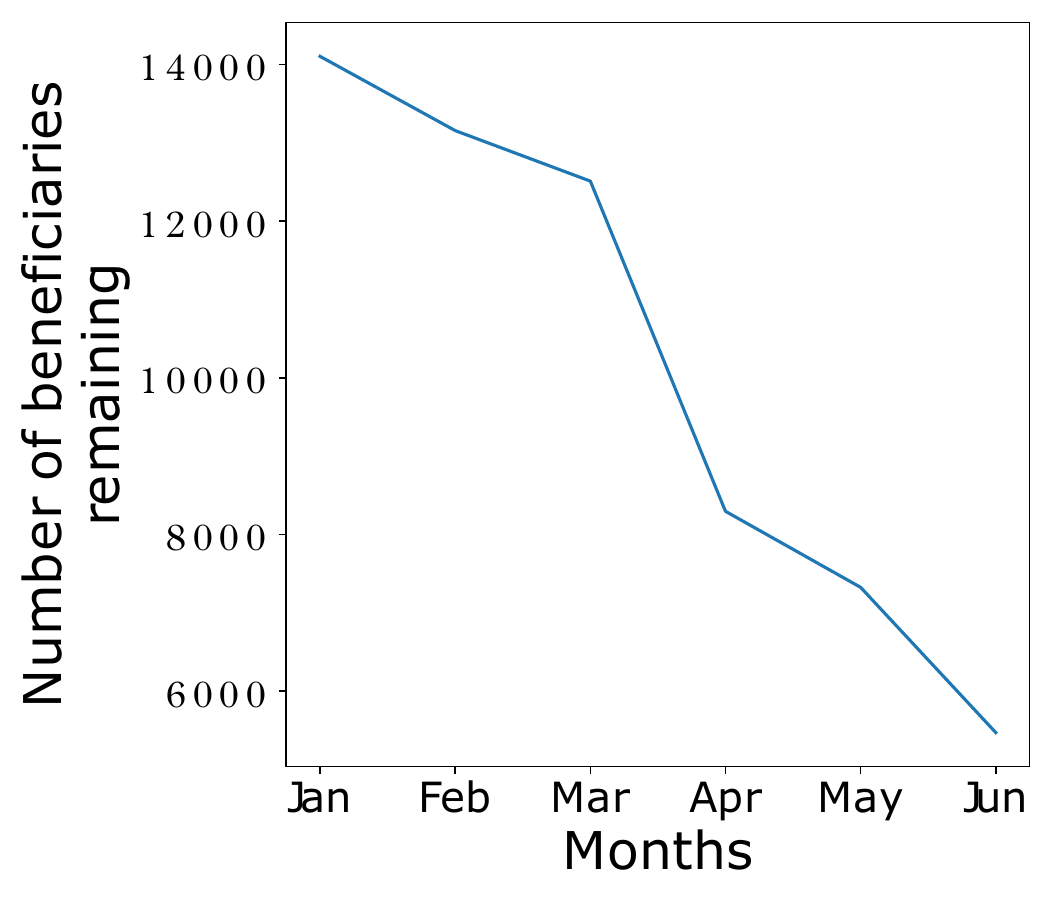}} 
    \end{subfigure}%
    \caption{(a) Around 23\% beneficiaries are unreached despite 9 attempts 
    (b) Number of beneficiaries (from the set of those enrolled in January 2022) remaining in the program after each month  }
    \label{fig:challenges}
\end{figure}

Despite multiple call attempts, our secondary analysis (see Fig. \ref{fig:challenges} (a)) shows that approximately 23\% of beneficiaries are not reached on average. Moreover, there is evidence of beneficiaries having preferences over the call time-slots,
which could contribute to reduced listenership and adherence~\cite{bashingwa2021assessing}. 
These are crucial challenges to address since beneficiaries who listen to less than 25\% of the AVM for 6 weeks in a row, are automatically dropped out of the program -- 
losing access to critical health information. Secondary analysis also shows that more than 50\% of beneficiaries enrolled in Kilkari in January 2022 in Odisha were dropped off within 6 months (Fig. \ref{fig:challenges} (b)). 
It is important to note that new beneficiaries are constantly being enrolled in the system, and hence 
AVM delivery
bandwidth does not go unutilized.
This motivates the need for optimizing preferred time slots, especially for those whose listenership is mainly impacted due to their availability to answer calls. Besides AVM time slot optimization, interventions continue to assist and retain beneficiaries in the system, and are of particular importance to beneficiaries who need encouragement and support.
\sysname{} ensures that interventions are directed towards beneficiaries who stand to gain the most, by optimizing both AVM time slots and intervention planning.

\subsection*{AVM Time Slot Optimization}

Optimizing call slots for AVMs is essential to improve engagement. 
Currently, Kilkari uses a random calling approach, attempting to reach beneficiaries at seven different timeslots throughout the day. Instead, \sysname{} optimizes AVM calls to maximize the likelihood of pick up based on past behaviours.

\subsection*{Intervention Planning}
\label{Intplan}

\sysname{} has two possible interventions available - ASHA visit, 
and an automated phone call reminder. Given the past history of listenership, beneficiaries are selected for an intervention based on the urgency, i.e, how long they are predicted to remain in an engaging state, to avoid drop-offs as well as boost listenership.

\textbf{ASHA visit scheduling}:
An ASHA (Accredited Social Health Activist) is a trained female community health worker 
who raises awareness about healthcare and bridges the gap between the community and the public health system. Among their many duties, ASHAs also help bring attention to the Kilkari program and register pregnant women or young mothers in their area into the program. In cases when a beneficiary is predicted to be at a risk of dropout, an ASHA can visit the beneficiary's home to discuss the importance of Kilkari in their pregnancy journey and ways to resolve any potential hindrances to their listenership. 
However, due to the limited bandwidth of the relatively few ASHA workers per district, 
there is a limit on how many beneficiaries can be reached. 
In order to test our system in a demanding, low-resource scenario, we opted for a $1\%$ (of the total number of beneficiaries) budget for the number of ASHA interventions. This number was selected as a realistic testcase in deliberation with our partner, ARMMAN.

\textbf{Automated Voice Call reminders}: 
The automated intervention calls highlight the importance of Kilkari and remind the beneficiaries to keep listening.
Similar to the ASHA interventions, we also selected an $1\%$ (of the total number of beneficiaries) budget for the calls.

\section*{Data Description}
ARMMAN stores anonymized call logs of beneficiaries' listenership of Kilkari AVM's. 
These logs contain information such as duration of call listened to, number of attempts to reach the beneficiary and the date and time of call attempts.

\subsection*{Pickup Rates}
To model the time slot planning algorithm, we estimate the pick-up rate from anonymized call logs data of $4000$ beneficiaries who were registered in Odisha and who had received a call in the first week of 2022. We consider $5$ time slots of two hours each from 8AM to 6PM plus one additional slot each for before 8AM and after 6PM, resulting in 7 time slots total.
The pickup rate for time slot \textit{x} is then estimated as the ratio between the number of calls picked up, divided by the number of calls made (in that timeslot).

\subsection*{Intervention Campaigns}
Using anonymised call logs, we perform secondary analysis of awareness campaigns conducted by ARMMAN in June 2023 
in 6 blocks in the Indian state of Odisha. 
In 3 blocks, the ASHA health workers were informed to visit randomly selected beneficiaries and encourage them to listen to the AVMs. To confirm that the visit happened, beneficiaries would send a missed call to a predefined phone number, which helped identify the date of the ASHA visit. In 3 other blocks, beneficiaries were sent an automated voice call reminder to encourage listening to the scheduled AVMs. Our secondary analysis also considers 6 other blocks with similar demographics but without any campaigns conducted. 
For the ASHA campaign, there were $148$ missed calls, while $1519$ Automated Voice Call Reminders were answered.
The automated call logs provide the listenership of these beneficiaries before and after the intervention.
These can be compared with the listenership trend of $3078$ beneficiaries from the blocks which didn't receive any interventions. 

We observe a $16\%$ increase in listenership through ASHA visits and $5\%$ 
increase in listenership from CALL campaigns (across all intervened beneficiaries). ASHA visits are, as expected, more effective because of the personal interaction with the beneficiaries.
We use this real data to train the proposed system to learn an optimal allocation of intervention resources, and to evaluate it in simulation to show its potential.

\section*{Time Slot Scheduling}

As mentioned, currently Kilkari employs a random calling approach, reaching out to beneficiaries at one of seven different timeslots throughout the day. Under this "retry" scheme, the system makes a maximum of $R$ call attempts per week until the beneficiary picks up, where currently $R=9$. However, the program does not remember which timeslot worked best for reaching the beneficiary. Consequently, there is no regular anticipation of the call timing, leading to potential wastage of a large number of retries. 

To address this issue and reduce the number of retries, we propose implementing a multi-armed bandit model -- specifically Upper Confidence Bound (UCB) -- for each beneficiary. Here, each "arm" represents a specific timeslot. "Pulling the arm" corresponding to timeslot $t$, implies sending the AVM at timeslot $t$. After each AVM call attempt, we record whether the call was answered or not.
\sysname{} maintains independently
its own estimates on the empirical mean and upper confidence bounds of each arm (beneficiary).
Let $\mu_{i,j}$ denote the empirical mean payoff of slot $j$ for beneficiary $i$.
Let $\nu_{i,j}$ denote the number of times slot $j$ was selected by the beneficiary $i$ so far. Since \sysname{} will run the algorithm independently for each beneficiary, for simplicity we can drop the subscript $i$. The number of pulls (or AVMs to beneficiary) each week is determined by the number of retries $R$, pulling up to $R$ slots until a call is successfully answered. 
The timeslot ($a_\tau$ at timestep $\tau$) is chosen based on the following UCB equation:
$$a_\tau = \arg\max_j \mu_j + \sqrt{2 (\log \tau) / \nu_j}$$

\section*{Intervention Planning} \label{sec: Intervention Planning}

In a RMAB setting, a planner can act (intervene) on $k$ out of $N$ arms (beneficiaries) at each timestep (due to budget constraints). Each arm is associated with a state, which transitions according to a Markovian transition function at each time-step. Several RMAB based solutions have been successfully developed 
~\cite{mate2022field,verma2022saheli,wang2023scalable,hsu2018age,kadota2016minimizing}, all of which, assume that transition probabilities follow the Markov property. 
Recent work, though, has demonstrated the importance of incorporating history~\cite{danassis2023TARI}, to capture \emph{temporal dependencies} in human behaviour. 
Unlike~\cite{verma2022saheli}, Kilkari also lacks the rich demographic features and hence \sysname{} relies on the use of call history data. Moreover, the section below validates that beneficiaries' behaviour in the Kilkari program is indeed history-dependent (non-Markovian). Therefore, we propose a non-markovian Time-Series Bandit model for multi-action setting to allocate health resources in Kilkari.

\subsection*{Non-Markovian Behaviour in Kilkari} \label{sec: Non-Markovian Behaviour in Maternal Healthcare Data}

In this section we 
demonstrate the benefit for a non-Markovian (or higher order Markov) model in explaining the beneficiaries' behaviour in Kilkari. For an $h$-order Markov process, the next state depends on the preceding $h$ states. More formally, let $x_1, \dots, x_t$ be the elements of the process. For a Markov process of order $h$, $(X_t)_{t = 1}^\infty$, it holds that $\forall t$, $\Pr[X_{t} = x_{t} \mid x_{t-1}, x_{t-2}, \ldots,  x_{1}] = \Pr[X_{t} = x_{t} \mid x_{t-1}, x_{t-2}, \ldots, x_{t-h}]$.
\noindent
Hence $h=1$ represents a traditional MDP, while for $h>1$, an $h$-order Markov process can be viewed as a first order Markov process on an expanded state space $s' = \times_{i=1}^h s_i$.

\begin{figure}[t!]
    \centering
    \includegraphics[width=0.7\linewidth]{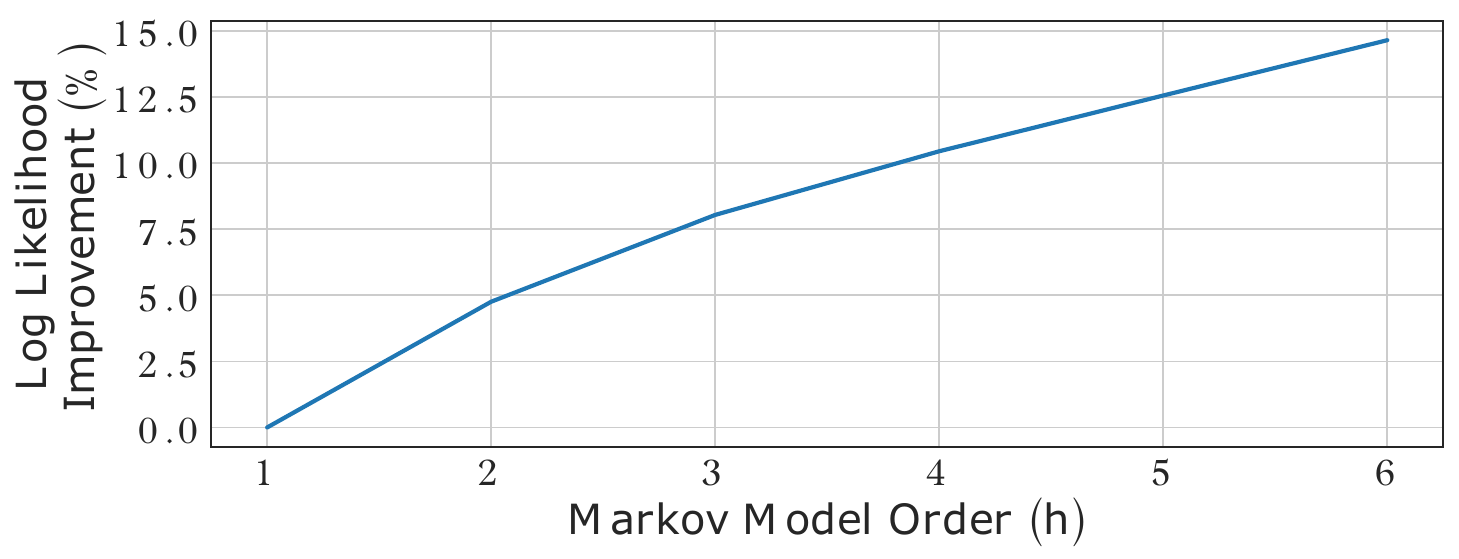}
    \caption{Relative (with respect to $h=1$) improvement in log likelihood. The probability of observing the historical trajectories in our dataset increases as we increase the order $h$ of the underlying Markov model. This suggests non-Markovian behaviour for the beneficiaries in Kilkari.}
    \label{fig: log_likelihood_improvement}
\end{figure}

As a measure of how closely the model can explain the beneficiaries' behaviour, we compute the log-liklihood of observing the historical trajectories of each beneficiary on Kilkari assuming an underlying MDP of order $h$. 
This is easily done by calculating empirical transition probabilities (assuming MDP of order $h$), and then computing the log-likelihood of each trajectory. $l(h) \triangleq -\log \mathcal{L}(h \mid x) \triangleq -\log \Pr (X=x \mid \text{model of order $h$})$. Finally, we average over all trajectories. Figure \ref{fig: log_likelihood_improvement} shows the relative improvement $-\left(\frac{l(h) - l(h=1)}{l(h=1)}\right)$ in negative log-likelihood for order $h$ Markov processes, 
with up to $15\%$ improvement for $h=6$.

\subsection*{Multi-Action Time Series Bandits}

We model the intervention scheduling as a Multi-action TSB problem $\{\mathcal{N}, \mathcal{K}, (X^{n \in N}_t)_{t=1}^H, R\}$, where $\mathcal{N}$ is the set of \emph{independent} arms, $\mathcal{K}$ is a set of budget constraints -- one per type of action $i$ -- 
$(X^{n \in \mathcal{N}}_t)_{t=1}^H$ is the associated transition process for arm $n$ for time horizon $H$, and $R: (X_t)_{t=1}^H \rightarrow \Re$ is the reward function.
The goal is to maximize the cumulative reward over the time horizon $H$. In our problem setting, the reward for a beneficiary at every timestep is the duration of call listened to in a week.

Each arm $n$ follows a history-dependent (i.e., non-Markovian) transition process. Let $s_{n, t} \in \mathcal{S}$, and $a_{i, n,t} \in \mathcal{A}$ denote the state and action taken on arm $n$, respectively, in timestep $t$. States are continuous in $[0, 1]$, and represent the `level of engagement' of a beneficiary, with higher numbers representing a higher level of listenership in seconds. States are fully observable. The action set consists of three actions:
(i) a visit by an ASHA worker, (ii) an automated voice call
or (iii) a no-cost passive action.

\subsection*{MA-TARI: A New Arm Ranking Index}

We follow a two stage process for planning which arms to choose for a particular action (intervention). First, we learn a model of beneficiaries' future listenership based on past behaviour through a Time-Series Forecasting (TSF) model. Next for each arm $n$ independently, we calculate two quantities recursively using our TSF model for each intervention. (i) The number of weeks $u_{n,i}$ it will take for beneficiary $n$ to fall below a threshold of listenership if we act once (using one of our intervention actions) on arm $n$ using intervention $i$ and never act again; (ii) The number of weeks $v_n$ it will take for beneficiary $n$ to fall below a threshold of listenership if we never act (passive action at each timestep). Then the Multi-Action Time-series Arm Ranking Index (MA-TARI) for arm $n$ and intervention $i$ is given by:
$$m_{n,i} = u_{n,i}/v_n$$ 

Since we independently compute one index per intervention, hereafter we will use TARI (instead of MA-TARI) when we refer to the index of a specific intervention.
To optimize the allocation of interventions to beneficiaries given the budget per intervention, we propose and compare four policies for the multi-action setting:
\begin{enumerate}

\item \textbf{MA-TARI Integer Linear Programming} :  We solve an ILP to find the optimal action $i$ for every arm $n$, such that we maximise the sum of MA-TAR indices $m_{n,i}$ (i.e., cumulative expected benefit from the interventions) while respecting the budget (first constraint) and making sure each beneficiary will only receive one intervention (second constraint). Intuitively, maximizing the sum of MA-TARI, optimizes the number of weeks the beneficiaries remain engaging with the program (based on our TSF model predictions), and thus the total listenership as well. We solve the following program (where $y_{n,i}$ denote the selection variables):
\small
\begin{eqnarray*} \label{Eq: tari ILP}
  \max_{y_{n,i} \in \{0,1\}} \ \ & & \sum_{n, i} y_{n,i} m_{n,i} \\
  s.t. \ \ & & \sum_n y_{n,i} \leq k_i, \forall i \nonumber \\
  & & \sum_i y_{n,i} \leq 1, \forall n \nonumber
\end{eqnarray*}
\normalsize

\item \textbf{MA-TARI Greedy}: Solving an LP for a large number of arms can be computationally challenging. As such, we also propose a greedy approach, which is more computationally tractable, especially for a low resource NGO. In the greedy policy, for each intervention $i$, we select the top $k_i$ arms with the highest TAR index without considering 
other interventions. If an arm is selected by two separate interventions, then we favor the one with the higher index. As demonstrated in the Experiments and Results Section, in this context, the computationally tractable greedy approach performs on par with the ILP.

\item \textbf{Random} : Randomly choosing $k_i$ beneficiaries per intervention $i$.

\item  \textbf{Control} : No interventions were given to this group.
    
\end{enumerate}

Finally, note that while we focus on Kilkari as our application domain, \emph{the proposed MA-TARI is potentially applicable to any multi-action, non-Markovian RMAB problem}.

\section*{Experiments and Results}

\subsection*{Time Slot Planning}

\begin{figure}[ht]
    \centering
    \includegraphics[width=0.8\columnwidth]{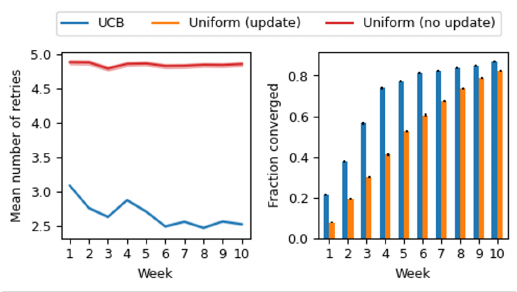}
\caption{[Left] Average number of calls performed each week for each beneficiary until a call is successfully picked up. [Right] Fraction of beneficiaries for which the learned pickup rate of the best slot has converged (within a difference of $\leq 0.15$ from the true mean).}
   
\label{fig:bandits-results}
\end{figure}

Figure \ref{fig:bandits-results} evaluates the effectiveness of UCB in learning an optimised time slot for each beneficiary
while adhering to the retry scheme $R=9$. We average over 30 runs for each beneficiary, and over 4000 beneficiaries.

Figure \ref{fig:bandits-results} [left] illustrates how UCB effectively reduces the number of calls needed to reach a beneficiary each week. A uniformly random algorithm that has no update on the timeslots' pickup rates -- denoted \textit{Uniform (no update))} -- requires an average of 4.86 calls per week for a successful call, while UCB accomplishes the same results by cutting the number of retries in almost \emph{half}, using as few as 2.5 calls per week by week 6. 

To further illustrate the effectiveness of UCB, we also compare it with \textit{Uniform (update)}, where we give the uniform random algorithm the additional ability to update its mean of call success rate per slot. Figure \ref{fig:bandits-results} [right] demonstrates the effectiveness of using UCB in learning the true pickup rate of the best timeslot. Within 4 weeks, UCB achieves a pickup rate difference of less than 0.15, compared to the true pickup rate, for 75\% of the population. In contrast, the Uniform (update) approach takes 8 weeks to reach a similar level of accuracy. 

\subsection*{Intervention Planning} \label{sec: Intervention Planning results}

The call logs provide us with beneficiaries listenership behaviour before and after the intervention (ASHA visit or CALL), and thereby each beneficiary's listenership trajectory as well as the action at every timestep.
\subsubsection{Time Series Forecasting Models} \label{sec: Time Series Forecasting Models}
We split the time series trajectories of beneficiaries into a 70-30 train-test split. We compared a range of time series forecasting models, such as LSTMs, BiLSTMS, Transformers, and adding time-based vector representations and attention layers to predict future listenership. Specifically, all these models take as input past states (duration of call listened to) and actions (ASHA intervention, CALL intervention, or no intervention) to predict listenership in the next timestep (output).

Comparing the performance of these models 
revealed that
all models have a similar Mean Absolute Error. We conducted this experiment over a range of context lengths, and observed similar results. On average, the maximum deviation of MAE of any model from the MAE of the LSTM model was less than 0.01. 
Thus we opted to use the simple LSTM architecture for computational efficiency as that would be preferred by our NGO partner which operates in a low resource environment. Our chosen model achieved MAE of $0.23$ on average for one step prediction on real data. We used a history of $h=4$ weeks for training and evaluation.

\subsubsection{Multi-Action TAR}
We use the TSF model to compute the MA-TAR index. In order to test its efficacy, we use a ground-truth behaviour simulator which simulates the true-listenership given the action suggested by the policy and past listenership. This ground-truth simulator is also a Time-Series Forecast model but is fitted on the entire dataset.

\begin{figure}
    \centering
    \includegraphics[width=0.9\linewidth]{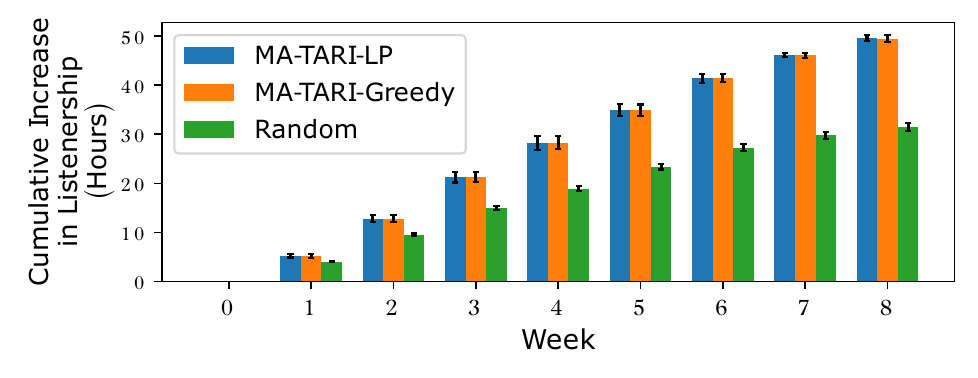}
    \caption{Comparison of different policies for planning multi-action interventions as compared to Control group.}
    \label{fig:policy_comparison}
\end{figure}

\begin{figure}[t!]
    \centering
    \begin{subfigure}[c]{0.5\columnwidth}
    \includegraphics[width=1\linewidth]{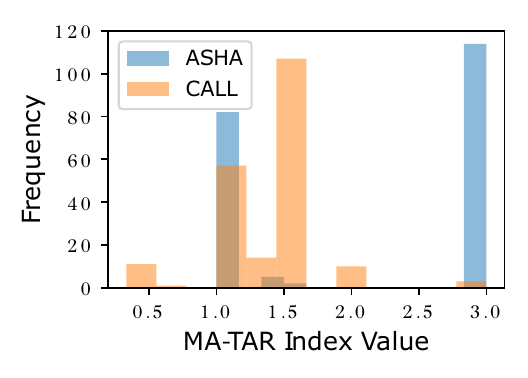}
    \caption{ }
    \label{fig:distribution}
    \end{subfigure}\hfill
    \begin{subfigure}[c]{0.5\columnwidth}
    \centering
    \includegraphics[width=1\linewidth]{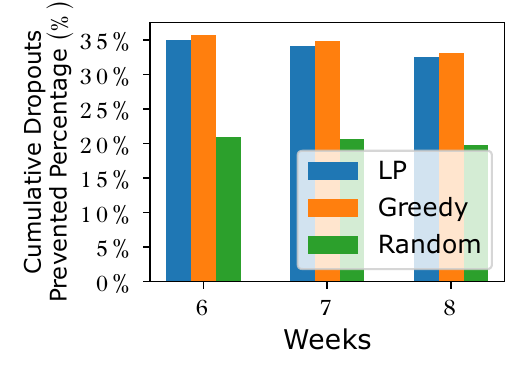}
    \caption{ }
    \label{fig:dropouts_graph}
    \end{subfigure}\hfill
    \caption{[Left] Distribution of Multi-Action TAR Indices for beneficiaries that receive interventions. [Right] Prevention of dropouts compared to control.} 
    \label{fig:distribution and dropouts_graph}
\end{figure}

In Fig. \ref{fig:policy_comparison}, we show how the two MA-TARI policies and the Random policy perform with respect to the control policy. Specifically, we measure the cumulative increase in listenership (in hours) every week, over the control group, for up to 8 weeks. In the simulation, we consider 4000 beneficiaries and allow 1\% ASHA interventions and 1\% CALL interventions every week. Fig. \ref{fig:policy_comparison} shows that the interventions significantly help in improving listenership over time. In particular, the MA-TARI policy results in 49.6h and 49.47h 
of additional content listened by the beneficiaries by the 8th week (for the LP and Greedy variants, respectively). Compared to the 31.5h increase with the random policy, this corresponds to an almost 57\% increase in exposure to content. 
Moreover, Fig.~\ref{fig:distribution} shows the distribution of the MA-TAR indices for the ASHA visit and call actions, highlighting the potential to delay the poor listenership state by 3x weeks in the best case. As expected, the ASHA intervention is more effective than the call reminder. 

Another important metric is identifying and proactively preventing drop-outs from the program. Fig. \ref{fig:dropouts_graph} depicts the percentage of drop-outs prevented (cumulatively), compared to the control group. Again, MA-TARI -- both the greedy and LP variants -- achieve a significant improvement in preventing drop-outs (33\% at the last week). 

These results directly translate to real-world impact, as listening to content improves health literacy, which ultimately leads to better health outcomes for the mothers and their children. Importantly, note also that the greedy variant performs on par to the cumbersome ILP formulation, thus providing a computationally efficient, and scalable policy for a low resource organization such as our NGO partner.

\section*{Data Usage and Ethics}
Acknowledging the responsibility associated with real-world AI systems for undeserved communities, we have closely coordinated with domain experts from our partner NGO, ARMMAN, throughout our analysis. This study falls into the category of secondary analysis of the aforementioned anonymized dataset. There is no demographic information available. We use the previously collected engagement trajectories of different beneficiaries participating in the Kilkari program to train the predictive model and evaluate the performance.
All the data collected through the program is owned by the NGO and only the NGO is allowed to share data. The research group only accesses an anonymized version of the data.

\textbf{Bias and fairness}
There is no demographic data available for Kilkari beneficiaries. Nonetheless, prior studies such as \cite{mohan2021can}
point out that exposure to Kilkari helps improve health behaviors among the most marginalised, thus helping to close the gap across some inequities in the population. They also indicate that the more marginalised population benefits from higher number of retries in Kilkari calls. 
\sysname{} potentially helps reduce inequities by  reducing the number of retries required, and by improving listening of Kilkari messages particularly amongst low listeners.

\section*{Path to Deployment}

\sysname{} is intended to be deployed at a national scale in India.
To that end, the next step
will involve a cohort study in one state in India to validate \sysname{}'s benefits in the field. Following a trial deployment, the program will then be gradually scaled up. Of course we expect such an endeavour to present many new challenges, such as regional differences in listenership patterns, or local constraints on the number and types of interventions. We aim to draw from our experience in deploying SAHELI~\cite{verma2022saheli} -- the system for ARMMAN's smaller scale maternal mHealth program, mMitra -- in order to tackle the challenges that will arise. Most importantly, though, just as with \sysname{}'s development, all of the steps will be done in \emph{close collaboration with our partner ARMMAN}; with ARMMAN ultimately in charge of the actual deployment. 

\section*{Conclusion}
In this work, we present \sysname{}, a novel AI system designed to improve
listenership and adherence in Kilkari, the largest maternal mHealth program in the world, with 3 million active beneficiaries at any given time.
Key innovations in \sysname{} include a novel multi-action, non-Markovian Time Series Bandit model to optimise the allocation of interventions across  beneficiaries to improve engagement.
We also presented a bandit algorithm for identifying preferred time slots of beneficiaries. With real world anonymized call data from the Indian state of Odisha, we show the benefits of our novel AI techniques in terms of improved listenership and 
identification of preferred beneficiary timeslots.

\section*{Acknowledgments} \label{sec: Acknowledgments}
This material is based upon work supported by the National Science Foundation under Grant No. IIS-2229881.  Any opinions, findings, and conclusions or recommendations expressed in this material are those of the author(s) and do not necessarily reflect the views of the National Science Foundation.

\bibliography{aaai22}

\end{document}